\documentclass[useAMS,usenatbib,usegraphicx]{mn2e}
\usepackage{amsmath,amsfonts,amssymb}

\newcommand{\lya}{Ly$\alpha$}
\newcommand{\di}{\textrm{d}}

\newcommand{\1}{^{-1}}
\newcommand{\hm}{$h^{-1}$}
\newcommand{\vir}{_{\textrm{vir}}}
\newcommand{\bj}{$b_{\rm J}$}
\newcommand{\oxy}{12+\log_{10}\left[O/H\right]_{\sun}}

\title[Recycling]{The recycling of gas and metals in galaxy formation:
predictions of a dynamical feedback model}

\author[Bertone et al.]
{Serena Bertone$^{1}$\thanks{E-mail: s.bertone@sussex.ac.uk},
 Gabriella De Lucia$^{2}$ and Peter A. Thomas$^{1}$ \\
$^{1}$Astronomy Centre, University of Sussex, Falmer, Brighton BN1 9QH, United Kingdom \\
$^{2}$Max Planck Institut f\"ur Astrophysik, Karl Schwarzschild Str. 1,
85741 Garching bei M\"unchen, Germany}

\begin{document}

\date{Accepted April 2007}

\pagerange{\pageref{firstpage}--\pageref{lastpage}} \pubyear{2005}

\maketitle

\label{firstpage}

\begin{abstract}

We present results of a new feedback scheme implemented in the Munich galaxy
formation model. The new scheme includes a dynamical treatment of galactic
winds powered by supernovae explosions and stellar winds in a cosmological context.
We find that such a scheme is a good alternative to empirically--motivated recipes for feedback in galaxy formation.
Model results are in good agreement with the observed luminosity functions and stellar mass function for galaxies in the local Universe. In particular, the new scheme predicts a number density of dwarfs that is lower than in previous models. This is a consequence of a new feature of the model, that allows an estimate of the amount of mass and metals that haloes can permanently deposit into the IGM. This loss of material leads to the suppression of star formation in small haloes and therefore to the decrease in the number density of dwarf galaxies.
The model is able to reproduce the observed mass--stellar metallicity and luminosity--gas metallicity relationships. This demonstrates that our scheme provides a significant improvement in the treatment of the feedback in dwarf galaxies.
Despite these successes, our model does not reproduce the observed bimodality in galaxy colours and predicts a larger number of bright galaxies than observed.
Finally, we investigate the efficiency of metal injection in winds and in the intergalactic medium. We find that galaxies that reside in haloes with
$M\vir < 10^{12}$ \hm M$_{\sun}$ may deposit most of their metal mass into the
intergalactic medium, while groups and clusters at $z=0$ have lost at most a few percent of their metals before the bulk of the halo mass was accreted.
\end{abstract}


\section{Introduction}
\label{intro}

Dwarf galaxies are the most abundant galactic objects in the universe and play
a key role in galaxy formation.  Because of their shallow gravitational
potentials, feedback from star formation is most efficient in these galaxies
and is believed to successfully eject a large amount of their gas and metals
(e.g. \citealt{larson1974}; \citealt{dekel1986}; \citealt{maclow1999};
\citealt{sh2003}; \citealt{veilleux2005} and references therein).  As a
consequence, dwarf galaxies may have a fundamental role in polluting the
diffuse intergalactic medium (IGM, hereafter) with the products of stellar
nucleosynthesis (e.g. \citealt{aguirre2001}; \citealt{theuns2001};
\citealt{sh2003}; \citealt{serena}; \citealt{oppenheimer2006}).

Observations of galaxies and their gas content in the local universe indicate
that feedback in the form of galactic winds plays an important role in
regulating star formation (\citealt{garnett2002}; \citealt{kauffmann2003};
\citealt{tremonti2004}; \citealt{salzer2005}).  As mentioned above, the impact
of feedback is believed to be strongest in dwarfs.  It is however important to
note that feedback can be equally effective in temporarily reheating the
interstellar medium and quenching star formation in more massive galaxies
(\citealt{heckman2000}; \citealt{cecil2001}; \citealt{rupke2002};
\citealt{shapley2003}; \citealt{rupke2005}).  A variety of observational data
suggests that, for each unit of gas mass that is converted into stars, an equal
or larger amount of interstellar gas can be reheated and ejected from the disk
of a galaxy (\citealt{martin1999}; \citealt{pettini2002}; \citealt{rupke2005};
\citealt{martin2006}).  Since the ejected mass may be lost to the galaxy and
deposited into the IGM, this has a dramatic effect on the total amount of stars
that can be formed in a galaxy during its lifetime.

By regulating star formation, feedback also alters the properties of the entire
galaxy population. The shape of the luminosity function (i.e. the space density
of galaxies as a function of their luminosity) is believed to reflect the
effects of different feedback mechanisms. Different studies have demonstrated
that feedback by galactic winds is responsible for shaping the faint end of the
luminosity function (\citealt{white1978}; \citealt{dekel1986};
\citealt{kauffmann1993}; \citealt{somerville1999}; \citealt{benson2003}). More
recent work has focused on feedback from accreting black holes in active
galactic nuclei (AGN feedback, hereafter) as an important physical mechanism to
suppress the cooling flows in relatively massive haloes
(\citealt{dimatteo2005}; \citealt{croton2006}; \citealt{bower2006}).

While some of the gas (and metals) ejected by winds may fall back onto the
galaxy at a later time via galactic fountains or gravitational accretion, a
fraction of the ejected material may be deposited permanently into the IGM.
Metals have been detected in the lowest density regions of the universe probed
by the \lya\ forest in the spectra of high--redshift quasars, and the IGM
metallicity has been estimated to be in the range $10^{-3}-10^{-2} Z_{\sun}$
(\citealt{songaila1996}; \citealt{rauch1998}; \citealt{schaye2000};
\citealt{schaye2003}; \citealt{aguirre2004}; \citealt{simcoe2004}). Although
most of the intergalactic metals appear to be associated with galaxies
(\citealt{adelberger2005}; \citealt{stocke2006}), a significant fraction does
not show such a correlation \citep{pieri2006}.  Galaxies remain the
best candidates to explain the enrichment of regions at larger overdensities,
such as filaments, groups and clusters (\citealt{aguirre2001};
\citealt{theuns2002}; \citealt{delucia2004}; \citealt{serena};
\citealt{daigne2006}; \citealt{oppenheimer2006}).  As for the enrichment of the
lowest density regions, it is still not clear which is the contribution from
the first populations of stars (\citealt{murakami2005};
\citealt{matteucci2005}).

In this work, we implement a dynamical treatment of galactic winds in a galaxy
formation model in order to test if this leads to any improvement on previous empirically--motivated feedback schemes used in semi--analytic models. Such a scheme introduces a further level of sophistication in the treatment of feedback.
To this end, we use an updated version of the wind model of \citet{serena} and the Munich galaxy formation model used in \citet{delucia2007} (DLB07 hereafter). Our wind model entirely replaces the supernovae feedback scheme used in the semi-analytic model but it does not affect the AGN feedback, which does not depend on star formation \citep{croton2006}. We apply our model to the Millennium Simulation \citep{springel2005} and we compare our model results to those from the latest Millennium galaxy catalogue of DLB07\footnote{The catalogue is publicly available at:\\ http://www.mpa-garching.mpg.de/millennium/}.

This paper is organised as follows. In Section \ref{model} we describe the new
features of our feedback scheme and its implementation in the Millennium
simulation. Section \ref{abundances} shows results for the galaxy luminosity
functions, the stellar mass function, galaxy colours, and morphological
fractions at $z=0$. In Section \ref{chemic} we discuss the impact of winds on
the chemical evolution of galaxies. Finally, in Section \ref{enrich} we
investigate the role of winds in recycling gas and metals, and in enriching the
IGM through the cosmic time. We draw our conclusions in Section \ref{summary}.

\section{Galaxy formation}
\label{model}

In this work we use the Millennium
simulation\footnote{http://www.mpa-garching.mpg.de/galform/virgo/millennium} to
trace the evolution of dark matter haloes in a cubic box of 500 \hm Mpc on a
side \citep{springel2005}. The underlying cosmological model is a $\Lambda$CDM
cosmology consistent with a combined analysis of the 2dFGRS and first year WMAP
data \citep{spergel2003}. The cosmological parameters are $\Omega_{\rm
  m}=0.25$, $\Omega_{\rm b}=0.045$, $h=0.73$, $\Omega_\Lambda=0.75$, $n=1$, and
$\sigma_8=0.9$. The Hubble constant is parameterised as $H_0 = 100$ \hm km
s$\1$ Mpc$\1$.
The simulation follows $N=2160^3$ dark matter particles of mass $8.6\times 10^8$ \hm M$_{\sun}$. Since we require dark matter haloes to be formed by at least 20 particles, the minimum halo mass is $1.7\times 10^{10}$ \hm M$_{\sun}$, with a corresponding baryonic mass of about $3\times 10^{9}$ \hm M$_{\sun}$.

As basis for our model, we use the galaxy formation model of DLB07.  We refer
the reader to the original paper for more details about the different physical
processes explicitely modelled and their implementation.  We note that the
model used in DLB07 retains many prescriptions of the previous version of the
Munich galaxy formation models described in \citet{croton2006}, but differs in
a number of details.  DLB07 use a different construction of the dark matter
merger trees, modified as to effectively track the most massive progenitor
history. DLB07 adopt a Chabrier initial mass function \citep{chabrier2003} and
an improved treatment of dust attenuation which takes into account the
attenuation of young stars within their birth clouds.  In this work, we use the
same set of parameters adopted in DLB07.  The feedback parameters
$\epsilon_{\textrm{disk}}$, $\epsilon_{\textrm{halo}}$ and the reincorporation parameter $\gamma_{\textrm{ej}}$ (see Table 1 in \citet{croton2006} are not used in our model.  Our wind scheme also contains free parameters, which we discuss in the following and summarise in Table~\ref{params}.

\subsection{A dynamical feedback model}
\label{windmodel}

The feedback model used here is based on the work by \citet{serena}. Galactic
winds are treated as spherical blastwaves, powered by supernova explosions and
stellar winds, which expand in a cosmological context.  The physics of galactic winds is described by the equation of motion that governs their dynamical evolution.

The main difference between this analysis and that of \citet{serena} resides in
the fully self-consistent treatment of feedback.  While in \citet{serena} winds
are added ``on top'' of the semi--analytic prescriptions for galaxy formation,
here we integrate the dynamic treatment of winds together with the existing
prescriptions for star formation, chemical evolution and gas cooling.  This
results in a scheme in which the ejection and recycling of gas and metals is
dictated by the dynamic evolution of the winds, rather than by phenomenological
recipes (\citealt{kauffmann1993}; \citealt{somerville1999};
\citealt{kauffmann1999}; \citealt{springel2001}; \citealt{delucia2004};
\citealt{croton2006}).  An alternative, dynamically--motivated approach to
feedback in semi--analytic models has recently been presented by
\citet{monaco2004}.

We describe the dynamical evolution of a wind as a two-stage process: a
pressure--driven, adiabatic expansion, followed by a momentum--driven
snowplough. During the first stage, a spherical bubble of hot gas blows out the
disk of a galaxy and a wind is formed. The expansion of the bubble is driven by
the pressure of the outflowing gas and the equation of motion is expressed by
the conservation of the bubble energy $E_{\textrm{b}}$:
\begin{eqnarray}\label{energy}
   \frac{\di E_{\textrm{b}}}{\di t} & = & \frac{1}{2}\dot{M}_{\textrm{w}}
   v_{\textrm{w}} ^2 \left( R \right) - \frac{\di \Delta W}{\di t}
   + 4\pi R^2 \cdot \nonumber \\
   & & \left\{ \left[ \frac{1}{2}\rho_{\textrm{o}} v_{\textrm{o}} ^2 + 
   u_{\textrm{o}} - \rho_{\textrm{o}} \frac{GM_{\textrm{h}}}{R} \right] \left( 
   v_{\textrm{s}} - v_{\textrm{o}} \right) - v_{\textrm{o}} P_{\textrm{o}} 
   \right\} 
\end{eqnarray}
$R$ and $v_{\textrm{s}}$ are the radius and the velocity of the shock,
$\dot{M}_{\textrm{w}}$ and $v_{\textrm{w}}\left( R \right)$ the mass outflow
rate and the outflow velocity of the wind at the shock radius (see Subsec.
\ref{masses}). $\rho_{\textrm{o}}$, $P_{\textrm{o}}$, $u_{\textrm{o}}$ and
$v_{\textrm{o}}$ are the density, the pressure, the internal energy and the
outward velocity of the surrounding medium and $M_{\textrm{h}}$ the total mass
internal to the shock radius.  The term $\di \Delta W / \di t$ indicates the
rate at which gravitational energy is transferred from the swept--up gas to the
other components \citep{omk}.  The expansion of a wind is determined by the
balance between the forces acting on the wind: the injection of energy and
momentum from SN accelerates the expansion, while the combined effect of
gravity, thermal pressure, and ram pressure slows it down and might eventually
make the wind collapse.

When radiative losses become dominant and most of the energy transferred to the
swept--up gas is radiated away, the adiabatic stage is terminated and a thin
shell of cooled gas forms near the outer edge of the bubble. This typically
happens when the cooling time of the bubble becomes shorter than the age of the
wind. When the snowplough stage sets in, the expansion of the shell is
described by the conservation of the wind momentum:
\begin{eqnarray}\label{momentum}
   \frac{\di}{\di t}\left( m v_{\textrm{s}} \right)
   & = & \dot{M}_{\textrm{w}} \left( v_{\textrm{w}}\left( R \right) -
   v_{\textrm{s}} \right) - 
   \frac{GM_{\textrm{h}}}{R^2} m - \nonumber \\
   & & 4\pi R^2 \left[ P_{\textrm{o}} + \rho_{\textrm{o}} 
   v_{\textrm{o}} \left( v_{\textrm{s}} - v_{\textrm{o}}\right) \right],
\end{eqnarray}
where $m$ is the mass of the shell. 

The only formal difference between these equations and those used in
\citet{serena} consists in the removal of the free parameter $\varepsilon$.
This means that a wind expanding into a halo sweeps up all the hot gas it
encounters on its path, instead of just a fraction of it.  It is therefore
equivalent to assuming $\varepsilon=1$.  In our model winds are powered by all
galaxies in a halo and effectively blow ``from haloes'', receiving energy,
momentum and mass from all the galaxies in the halo (see sec.~\ref{ic}).  When
haloes merge and a former central galaxy becomes a satellite, its wind is
transferred to the central galaxy of the merged halo.  If the central galaxy is
already blowing a wind, the two are combined under the assumption that mass,
energy and momentum are conserved.  The mass and the metals ejected into the
IGM by the merged galaxies are added to the ejected component of the new
central galaxy.  

If the momentum of the outflowing material is not able to balance the momentum
losses by gravity and pressure forces, the wind can collapse and create a
galactic fountain. The dynamics of the collapsing wind can be traced by Eq.
(\ref{momentum}) until the shell falls back onto the central galaxy.  We
assume that all the mass in the collapsed wind is instantaneously
reincorporated into the hot component of the halo.  In previous
schemes \citep{delucia2004,croton2006} the ejected material is reincorporated
into the hot component, thereafter being again available for cooling, on a
time--scale that is proportional to the dynamical time--scale of the halo. \citet{delucia2004} argue that various observational results can be reproduced
only by models in which the ejected material is re-incorporated on a time-scale
comparable to the age of the Universe.  

If a large amount of energy and momentum is injected in a bubble, a wind can
overcome gravity and pressure forces and expand to a large radius. When the
wind reaches the Hubble flow, it escapes from the halo and its mass and metals
are lost for ever from the galaxy.  We define the transition between the halo
environment and the Hubble flow as the point where the velocity of the
surrounding medium, defined as $v_{\textrm{o}} = RH\left( z\right) -
v_{\textrm{esc}}\left( R \right) / 2$, is dominated by the Hubble flow term and
becomes positive.  Here, $H\left( z\right)$ is the Hubble constant at redshift
$z$ and $v_{\textrm{esc}} \left( R \right)$ is the escape velocity of the halo
at radius $R$, calculated assuming a NFW profile \citep{navarro1996}. A
consequence of this assumption is that the maximum radius to which winds can
travel is proportional to the virial mass of the halo. Winds outflowing from
small haloes escape at smaller radii than winds expanding out of larger haloes.
In the following, we will call ``mass in IGM'' the wind mass that reaches the Hubble flow.
This material cannot be re-incorporated and therefore does not partecipate
again in the evolution of the galaxy.

\subsection{Mass and metals in winds}
\label{masses}

The evolution and exchange of metals between the different galactic components
follows the same scheme proposed in \citet{delucia2004}, modified to integrate
self-consistently the removal of gas and metals from the cold gas. When stars
form, metals are produced and instantaneously returned to the interstellar
medium. Subsequently, they can be deposited into the hot gas and ejected
component by feedback processes.

During the adiabatic stage, the outflowing mass creates the over-pressurized
bubble of hot gas. The mass of the bubble $M_{\textrm{b}}$ is determined by the
conservation of mass law:
\begin{equation}\label{mass_e}
\frac{\di M_{\textrm{b}}}{\di t}=\dot{M}_{\textrm{w}} + 4\pi R^2 \rho_{\textrm{o}} \left( v_{\textrm{s}} - v_{\textrm{o}}\right),
\end{equation}
where the first term on the right hand side represents the mass of cold gas
reheated by SN explosions and stellar winds and the second term defines the
amount of hot gas swept up by the wind and accreted during the expansion in the
halo.  In a snowplough, most of the mass accumulates in a thin shell of mass
$m$:
\begin{equation}\label{mass_m}
\frac{\di m}{\di t}=\dot{M}_{\textrm{w}} \left( 1- \frac{v_{\textrm{s}}}{v_{\textrm{w}}\left( R \right)} \right) + 4\pi R^2 \rho_{\textrm{o}} \left( v_{\textrm{s}} - v_{\textrm{o}}\right).
\end{equation}
A fraction $\dot{M}_{\textrm{w}} v_{\textrm{s}} / v_{\textrm{w}}\left( R
\right)$ of the outflowing ISM does not reach the shell and remains in the
cavity behind the shock.  Assuming that the amount of metals exchanged is
proportional to the corresponding amount of mass (i.e. neglecting any
metal-loading effect), the corresponding equations for metals are:
\begin{equation}\label{mass_z}
\frac{\di M_{\textrm{bz}}}{\di t}=\dot{M}_{\textrm{w}} Z_{\textrm{cold}} + 4\pi R^2 \rho_{\textrm{o}} \left( v_{\textrm{s}} - v_{\textrm{o}}\right) Z_{\textrm{hot}},
\end{equation}
where $Z_{\textrm{cold}}$ is the metallicity of the cold gas and
$Z_{\textrm{hot}}$ the metallicity of the hot gas, and:
\begin{equation}\label{mass_zshell}
\frac{\di m_{\textrm{z}}}{\di t}=\dot{M}_{\textrm{w}} \left( 1- \frac{v_{\textrm{s}}}{v_{\textrm{w}}\left( R \right)} \right) Z_{\textrm{cold}} + 4\pi R^2 \rho_{\textrm{o}} \left( v_{\textrm{s}} - v_{\textrm{o}}\right) Z_{\textrm{hot}},
\end{equation}
A fraction $Z_{\textrm{cold}} \dot{M}_{\textrm{w}} v_{\textrm{s}} /
v_{\textrm{w}}\left( R \right)$ of the outflowing metals remains in the cavity
behind the shell.

Eqs. \ref{mass_e}--\ref{mass_zshell} imply that the amount of mass and metals ejected from the cold gas are $\dot{M}_{\textrm{w}} \di t$ and $\dot{M}_{\textrm{w}} Z_{\textrm{cold}} \di t$ respectively.
In our model, we assume that supernova ejecta efficiently mix with the ISM before leaving the disk of the galaxy, as done in \citet{delucia2004}.
The amount of mass and metals removed from the halo hot gas and swept up on to a wind are $4\pi R^2 \rho_{\textrm{o}} \left( v_{\textrm{s}} - v_{\textrm{o}}\right) \di t$ and $4\pi R^2 \rho_{\textrm{o}} \left( v_{\textrm{s}} - v_{\textrm{o}}\right) Z_{\textrm{hot}} \di t$  respectively. The removal of hot gas from the halo is in principle similar to the implementation of the ``ejected mass'' in the \citet{croton2006} model, but here it is justified on the basis of the dynamical evolution of the winds and their sweeping up the gas on the path of the expansion.

\subsection{Wind initial conditions and model parameters}
\label{ic}

Observations of galaxies in the local and high redshift universe suggest that
the mass loss rate of winds is roughly proportional to the star formation rate
(\citealt{martin1999}; \citealt{pettini2002}; \citealt{rupke2005};
\citealt{martin2006}).
A similar behaviour is predicted by the models of \citet{monaco2004} and \citet{shu}. Only a weak dependence of the mass loss rate on the virial velocity of haloes is instead observed (\citealt{martin1999}; \citealt{rupke2005b}).
In our model, we parametrise the mass loss rate
$\dot{M}_{\textrm{w}}$ as a function of the star formation rate
$\dot{M}_{\star}$ and of the virial velocity of the halo $v \vir$:
\begin{equation}\label{massloss}
\dot{M}_{\textrm{w}} = \eta  \dot{M}_{\star} \left( \frac{220\textrm{ km s}^{-1}}{v \vir} \right) ^{\alpha_{\textrm{w}}}.
\end{equation}
In the above equation, $\dot{M}_{\textrm{w}}$ represents the sum of the star
formation rates of all the galaxies in a halo.
The power law index $\alpha_{\textrm{w}}$ and the constant wind mass loss rate $\eta$ are free parameters.
According to this choice, all galaxies in a halo contribute to power the corresponding wind and to eject mass and metals.  The dependence of the wind mass loss rate on the halo virial velocity $v \vir ^{-\alpha_{\textrm{w}}}$ implies that the ratio $\dot{M}_{\textrm{w}} / \dot{M}_{\star}$ is highest in the smallest haloes.
We note that this definition of the mass loss rate is equivalent to that of the
reheated mass in \citet{delucia2004}. In our fiducial model we assume $\eta =2$ and $\alpha_{\textrm{w}} = 2$. 

At the galactic radius $R_{\textrm{gal}}$, where energy and momentum are
injected into the wind, the initial wind velocity $v_{\textrm{wo}}$ is calculated assuming that the
energy injection rate in winds $\dot{E}_{\textrm w} = \dot{M}_{\textrm{w}}
v_{\textrm{wo}}^2$ is constant and independent of halo properties. Given the dependence of the mass loss rate on $v \vir$ in Eq. \ref{massloss}, the wind initial velocity can be expressed as
\begin{equation}\label{windspeed}
v_{\textrm{wo}} = V_{\textrm{w}} \left( \frac{220\textrm{ km s}^{-1}}{v \vir}
\right) ^{-\alpha_{\textrm{w}} / 2},
\end{equation}
where the wind initial velocity $V_{\textrm{w}}$ is treated as a free parameter and we choose as fiducial value $V_{\textrm w} = 800$ km s$\1$. Table \ref{params} summarises the values of all the free parameters for our fiducial model.
The wind velocity at a radius $r$ is calculated as $v_{\textrm{w}}^2 \left( r \right) = v_{\textrm{wo}} ^2 - 2\left[ \phi\left( r\right) - \phi \left( R_{\textrm{gal}}\right)\right]$, where $\phi$ is the gravitational potential of the halo, calculated assuming an NFW profile.

\begin{table}
\begin{center}
\begin{tabular}{|c|c|}
\hline
Model parameter &  Fiducial value \\
\hline
$\eta$ & 2 \\
$V_{\textrm{w}}$ & 800 km s$\1$ \\
$\alpha_{\textrm{w}}$ & 2 \\
\hline
\end{tabular}
\caption{Summary of model parameters.}
\label{params}
\end{center}
\end{table}

Observations of galaxies both in the local and the high redshift universe
estimate wind velocities in the range 100--1500 km s$\1$
(\citealt{strickland2000}; \citealt{frye2002}; \citealt{shapley2003};
\citealt{rupke2005}; \citealt{rupke2005b}; see also \citealt{veilleux2005} for a complete review of the phenomenology of galactic winds). Since winds are multiphase phenomena, the detection of the outflowing gas in different wavebands gives information about wind gas components at different temperatures.  These may yield very diverse values for the wind velocity.  The X--ray emitting gas is believed to represent the most energetic phase of the wind, containing most of the metal--rich supernova ejecta. Simulations by \citet{strickland2000} demonstrate that this gas is outflowing with the highest velocities, but it is not representative of the bulk of the mass in winds. The latter - that is what our wind model is meant to describe - can be detected in less energetic wavebands and has much lower velocities (\citealt{rupke2005}; \citealt{heckman2000}).
\citet{monaco2004} implemented a feedback scheme in which the cold and the hot outflowing phases of a wind are followed individually. This scheme has been used by \citet{monaco2007} in their galaxy formation model {\small MORGANA}.

Although the large wind velocities given by Eq. \ref{windspeed} seeming to
indicate that winds can escape from large haloes, this never occurs in
practise. As we will show in the following, in haloes with $M \vir \gtrsim
10^{12}$ \hm M$_{\sun}$ the injected energy is never large enough to overcome the
gravitational potential and the pressure forces of the intra--cluster gas.  In
these haloes, the wind usually collapses within a few time-steps and its mass
and metals are reincorporated into the hot gas.

Our set of initial conditions differs from that used in \citet{serena}, who used the parametrisation for $v_{\textrm{wo}}$ and $\dot{M}_{\textrm{w}}$ proposed by \citet{shu}. In that paper, the wind mass loss rate and the wind initial velocity scale with the star formation rate as $\dot{M}_{\textrm{w}} \propto \dot{M}_{\star}^{-0.71}$ and $v_{\textrm{wo}} \propto \dot{M}_{\star}^{-0.15}$. The dependence of the wind initial velocity on the halo virial velocity in Eq. \ref{windspeed} can be expressed as $v_{\textrm{wo}} \sim 3.6v_{\vir}$. This is consistent with the parametrisation of \citet{oppenheimer2006}, who assume that $v_{\textrm{wo}} \sim 3v_{\vir}$, in agreement with the findings of \citet{martin2005}. \citet{oppenheimer2006} calculate the wind mass loss rate as $\dot{M}_{\textrm{w}} \propto v_{\vir}^{-1}$. This is equivalent to assuming $\alpha_{\textrm{w}}=1$ in Eq. \ref{massloss}. While our model assumes the same value of $\alpha_{\textrm{w}}$ in Eqs. \ref{massloss} and \ref{windspeed}, the model of \citet{oppenheimer2006} requires two different values of $\alpha_{\textrm{w}}$. This implies that the energy injection rate in winds $\dot{E}_{\textrm{in}}$ in the model of \citet{oppenheimer2006} scales with virial velocity as $\dot{E}_{\textrm{in}} \propto v_{\vir}$, while the momentum input $\dot{M}_{\textrm{w}} v_{\textrm{wo}}$ is independent of $v_{\vir}$, as predicted for momentum--driven winds by \citet{murray2005}. In our model we have assumed that the energy injection rate, and not the momentum injection rate, is independent of halo properties, as assumed for example by \citet{sh2003}.

\section{The abundance of galaxies}
\label{abundances}

In this Section we present results for the galaxy luminosity functions (LFs
hereafter, Subsec. \ref{galaxies}), the stellar mass function (Subsec.
\ref{massfunction}), and the morphology distribution of galaxies at $z=0$
(Subsec. \ref{morpho}). We discuss galaxy colours in Subsec. \ref{colour}.
Throughout, we highlight the role of winds in suppressing star formation in
faint galaxies, which is, perhaps, the most important achievement of our model.

\subsection{Galaxy luminosity functions}
\label{galaxies}

\begin{figure}
\centering
\includegraphics[width=8.4cm]{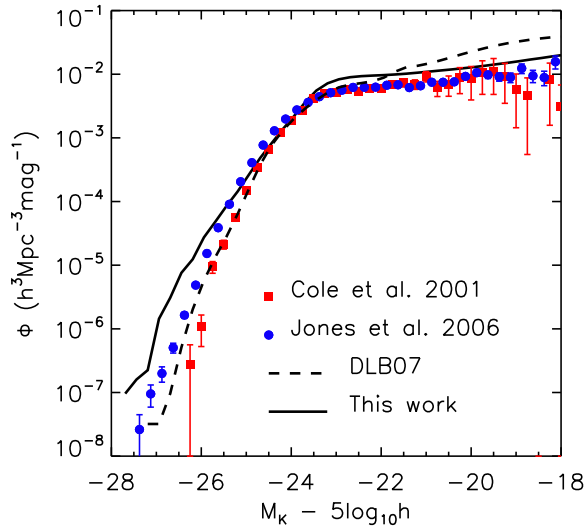}
\caption{K--band galaxy luminosity function at $z=0$. Results from this
  work (solid line) are compared to observational data from the 2dFGRS
  \citep{coles2001} and 6dFGS \citep{jones2006} surveys (filled squares and
  circles respectively). The dashed line shows the corresponding luminosity
  function from the model of DLB07.}
\label{kband}
\end{figure}

\begin{figure}
  \centering \includegraphics[width=8.4cm]{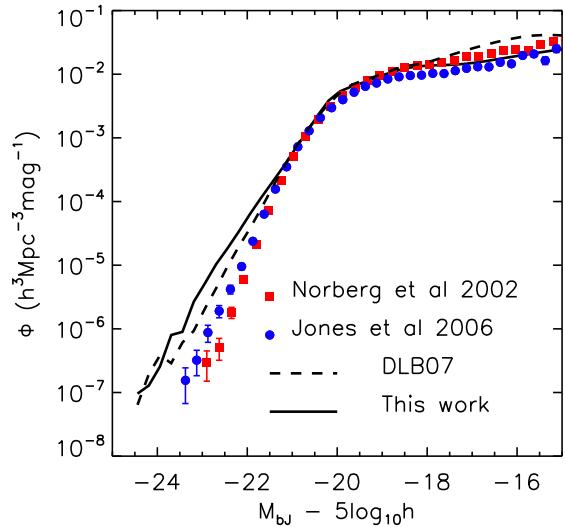}
\caption{As in Fig.~\ref{kband} but for the \bj--band galaxy luminosity
  function at $z=0$.}  
\label{bjband}
\end{figure}

Figs.~\ref{kband} and \ref{bjband} show the galaxy K--band and \bj--band LFs
from our model (solid line) and the corresponding observational data from the 2
degree Field Galaxy Redshift Survey (\citealt{coles2001}, 2dFGRS hereafter) and
the 6 degree Field Galaxy Survey (\citealt{jones2006}, 6dFGS hereafter).  In
both figures, the dashed line shows the corresponding luminosity function from
the model of DLB07. 

Our predictions differ from those of DLB07 both at the faint and at the bright
end of the LFs. Our model predicts a shallower faint end slope than DLB07,
bringing the abundance of the faint model galaxies closer to the observed
values.  Although the agreement with observational data is still not perfect,
it improves the results of DLB07 by a factor of two for galaxies fainter than
$M_{\rm K} > -22$.  The ability to accurately reproduce the slope of the faint
end of the galaxy LFs is peculiar to our model and is a consequence of our new
feedback scheme. In addition, the slope of the galaxy LFs does not appear to
be sensitive to the parameters regulating our feedback scheme.
DLB07 predict steeper slopes for the faint end of the \bj-- and K--band LFs, which are not seen in observations. We have found that our model can reproduce these features only when feedback is particularly inefficient in dwarfs (i.e. $\alpha_{\rm w}=0$, or $V_{\rm w}$ and $\eta$ have very low values).  In this case, the energy injected in winds is not enough to overcome the gravitational potential of even the smallest haloes, leading to a severe overestimation of the number of faint galaxies.

Figs.~\ref{kband} and \ref{bjband} also show that our model overpredicts the
abundance of bright galaxies both in the \bj -- and in the K--band.  This
overprediction is a consequence of the short reincorporation time-scales
obtained for relatively massive haloes. In our model, winds in
haloes with $M\vir \gtrsim 10^{12}$ \hm M$_{\sun}$ are unable to escape the
gravitational potential and typically collapse after a few million years. The
mass of the collapsing wind is instantaneously reincorporated into the hot gas and is again available for cooling, contributing to increase the luminosity of the central galaxy.  This problem could partly be solved by assuming a higher
efficiency for the AGN feedback (20--30 per cent higher than in
\citealt{croton2006}) or by increasing the reincorporation timescale of
collapsed winds in large haloes (as in \citealt{delucia2004}). However, since the aim of this work is to test the performance of a dynamical wind model in shaping the galaxy properties, assuming a non--instantaneous reincorporation time--scale defeats the main purpose of the model.

\subsection{Stellar mass function}
\label{massfunction}

\begin{figure}
\centering
\includegraphics[width=8.4cm]{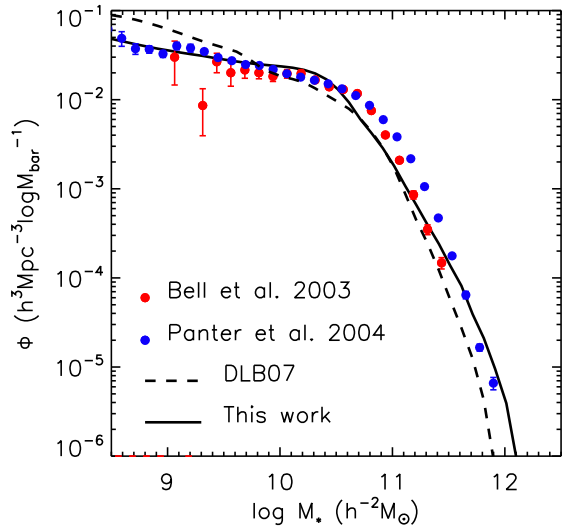}
\caption{Stellar mass function at $z=0$. Numerical results for our model and for DLB07 are compared to the observations of \citet{bell2003} and \citet{panter2004}.}
\label{mf}
\end{figure}

In this Subsection we focus on the stellar mass function of galaxies at $z=0$, which we show in Fig. \ref{mf}. Our results are compared to the observations of \citet{bell2003} and \citet{panter2004} and to the model of DLB07. We use the mass function results of \citet{bell2003} obtained from the SDSS $g$--band.
Our model shows a better agreement with the observed stellar mass functions than DLB07, and in particular it better reproduces the number density of faint galaxies and of the brightest galaxies. However, both our model and the model of DLB07 underpredict the spatial density of $M^{\ast}$ galaxies at the knee of the distribution, for $10^{10.5}$ \hm M$_{\sun} < M_{\star} < 10^{11.5}$ \hm M$_{\sun}$.

The stellar mass function should closely reflect the K--band LF. Although at first sight the lack of galaxies around $M^{\ast}$ does not seem to appear in the K--band LF (Fig. \ref{kband}), if we shifted the LF to lower luminosities by about 0.5 mag, we would recover exactly the same effect. This may be an indication that the galaxy luminosities are overestimated, or that, more likely, the galaxy formation model is unable to produce as many $M^{\ast}$ galaxies as there are in the real universe.
The discrepancy between our predictions for the K--band LF and for the stellar mass function could also be partly due to a difference between the mass--to--light ratios predicted by our model and those used by \citet{bell2003} and \citet{panter2004}. The high $\sigma_8$ value used in the Millennium simulation may further affect these results negatively.

We believe that the underestimation of the abundance of $M^{\ast}$ galaxies is due to a combination of effects. In this range of stellar masses, the effects of galaxy and AGN feedback overlap and as a result star formation may be suppressed more strongly than in other mass ranges. We find two possible ways to solve this problem. An overall increase in the efficiency of star formation or, alternatively, a decrease in the efficiency of feedback in galaxies as bright or brighter than $M^{\ast}$ helps to produce more galaxies at the knee, and in particular more early types. However, in the first case, much higher feedback is required to suppress the formation of dwarf galaxies and predict the correct slope of the faint end of the stellar mass function. On the other hand, reducing the efficiency of feedback in $M^{\ast}$ galaxies would imply that the efficiency to deposit supernova energy in winds depends critically on some unknown galaxy properties. In either case, there might be a connection between physical processes taking place at scales not resolved by current simulations and the efficiency of star formation and feedback that we have not been able to include in our modelling of galaxy formation.

\subsection{Galaxy morphologies}
\label{morpho}

\begin{figure}
\centering
\includegraphics[width=8.4cm]{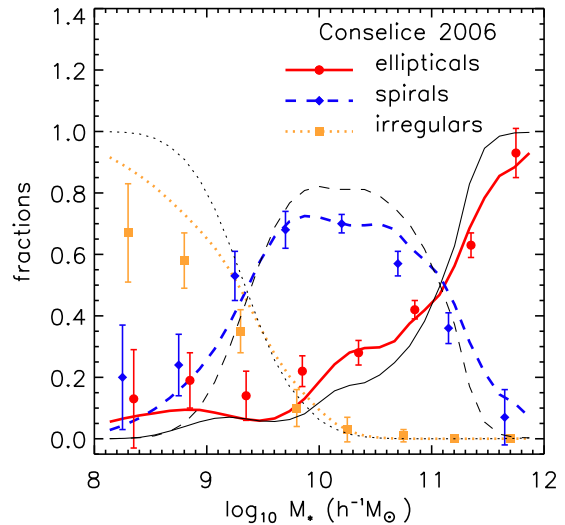}
\caption{Fraction of galaxy morphological types as a function of stellar mass. We define as ellipticals all galaxies with a dominant bulge component ($M_{\textrm{bulge}} / M_{\star} > 0.7$), as spirals all galaxies with $0< M_{\textrm{bulge}} / M_{\star} < 0.7$, and as irregulars all galaxies with no bulge. Model results are compared to the observational determination of \citet{conselice2006}. Thick coloured lines show the results of this work, while thin black lines are for the model of DLB07.} 
\label{fracs}
\end{figure}

In this section we analyse the morphological mix of our model galaxies at
$z=0$.  Fig. \ref{fracs} shows the fraction of galaxies with different
morphological types as a function of stellar mass.  Data points in
Fig. \ref{fracs} show observational determinations from \citet{conselice2006},
who analysed a sample of about 22000 galaxies with a visual morphological classification from the RC3 catalogue \citep{devaucouleurs}.  As a proxy for the morphology of our model galaxies,
we use the ratio between the bulge mass and the total stellar mass ($r =
M_{\textrm{bulge}} / M_{\star}$).  In particular, we classify as ellipticals
galaxies with more than 70 per cent of their stars in the bulge ($r>0.7$), as
irregulars all galaxies without a bulge ($r=0$), and as spirals all galaxies
with $0<r\leqslant 0.7$.  The threshold value $r=0.7$ is fixed arbitrarily, but
results do not depend significantly on it. In particular, independently of
the value of $r$, we find that ellipticals represent the dominant type at the
highest stellar masses, while the smallest galaxies are predominantly
irregulars.  Most of the galaxies with $10^{9.5}$ \hm M$_{\sun} < M_{\star} < 10^{11}$ \hm M$_{\sun}$ have a spiral morphology.  A lower value of $r$ would
slightly decrease the fraction of spirals at intermediate stellar masses, while
a higher value would slightly increase it.

Fig. \ref{fracs} shows that our model reproduces quite nicely the observational results. The main difference with the model predictions of DLB07 is found for the highest and lowest stellar mass galaxies. In the highest stellar mass bin
our model does predict a few spiral galaxies, in agreement with observations.
The model of DLB07 does not predict any spiral in this mass range, even
assuming a value of $r=0.6$.  This difference is due to the recycling of gas in
our feedback model, which is particularly efficient for galaxies in groups with
$M\vir \gtrsim 10^{13}$ \hm M$_{\sun}$. This results in galaxies with larger stellar masses to have larger disks in our model. We will discuss this feature of our model in more detail in Sec. \ref{enrich}. In the intermediate mass range, our model predicts a lower fraction of spirals than DLB07, in better agreement with observational results.  In the lowest mass bins, our model predicts a non--zero fraction of spiral and elliptical galaxies and a fraction of irregular galaxies lower than one.  We note, however, that the morphological classification of model galaxies in this mass range might not be robust, as they reside in haloes whose history can be followed back in time only for a few time--steps.

In our model, the mass of the bulge is determined by two processes: mergers and dynamical instabilities of the disk. As explained in detail in \citet{croton2006} and \citet{delucia2006}, the disk instability is treated as in \citet{mo1998} and depends on the density of stars in the galactic disk. The outcome of a galaxy merger depends on the baryonic mass ratio of the merging galaxies. Our feedback scheme modifies the mass of gas and stars in galaxies (e.g. see Fig. \ref{mf}) and so indirectly affects the morphological mix of model galaxies.

\subsection{Galaxy colours}
\label{colour}

\begin{figure}
\centering
\includegraphics[width=8.4cm]{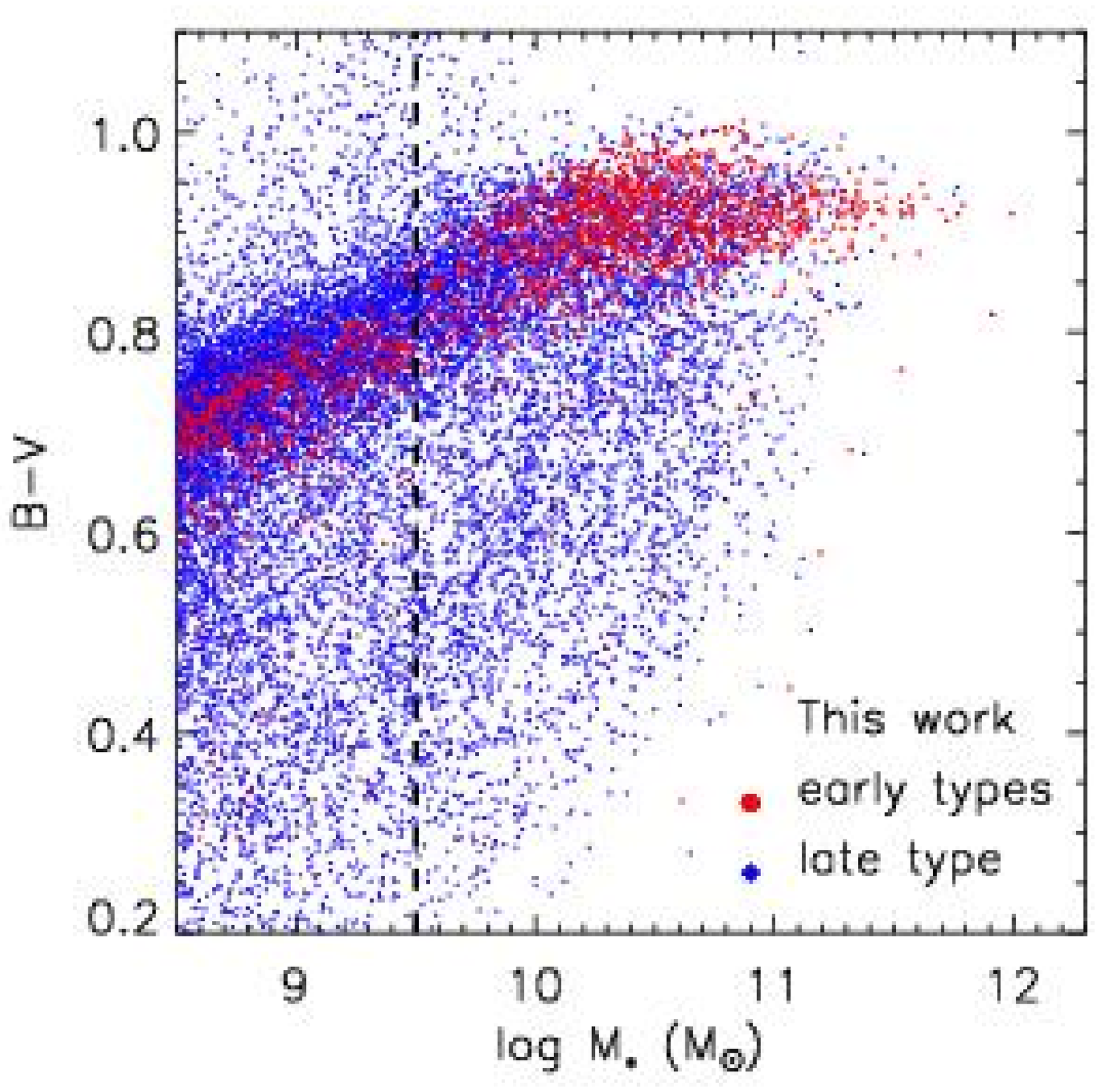}
\includegraphics[width=8.4cm]{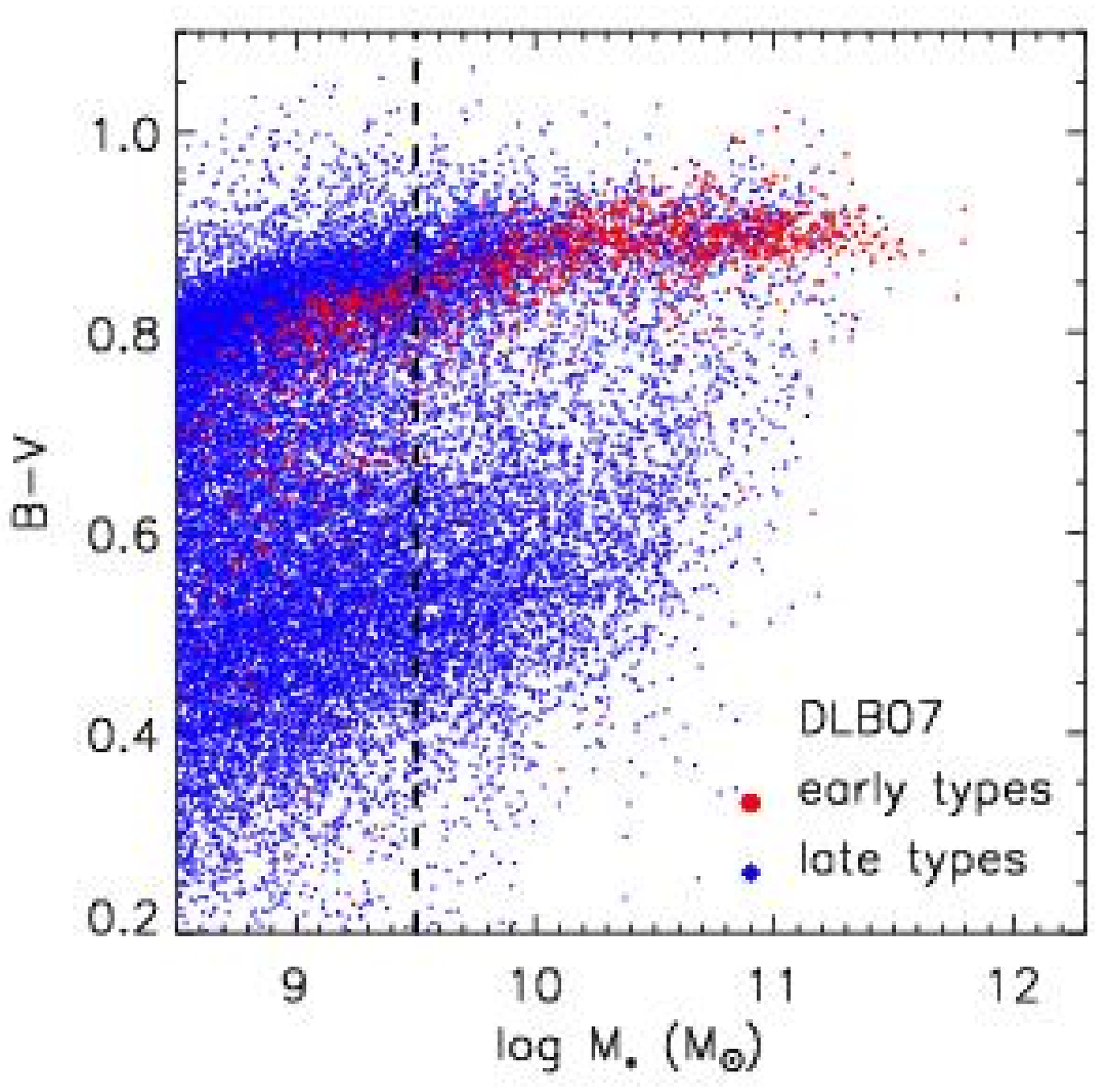}
\caption{The $B-V$ colours of model galaxies as a function of their stellar
mass.  The upper panel shows results from the model discussed in this paper
while the lower panel shows result from DLB07.  Galaxies are defined as
``late types'' and ``early types'' according to their bulge--to--total
luminosity ratio in B--band.  The vertical dashed line marks the limit where
the morphological classification of model galaxies can be considered robust.}
\label{color}
\end{figure}

In this Subsection we briefly discuss the colours of galaxies and the colour--magnitude relationship.
Fig.~\ref{color} shows the $B-V$ colour of model galaxies as a function of
stellar mass.  The upper panel shows results from the model discussed in this
paper while the lower panel shows results from DLB07.  Note that this figure
corresponds to Fig.~9 of \citet{croton2006}.  As in that work, the morphology
of model galaxies is here defined on the basis of their bulge--to--total
luminosity ratio in B--band.  The vertical line indicate the limit above which
the morphological classification can be considered robust. 

Fig.~\ref{color} indicates that there are several differences between our model
and the model of DLB07.  While DLB07 predicts a sharp colour bimodality, our
model shows a much smoother distribution of galaxy colours, with few blue
galaxies and most of the galaxies residing on a red sequence which appears
steeper and characterised by a larger scatter than the model results from
DLB07.  The suppression of the colour-bimodality is a consequence of the wind
model.  We have verified that most combinations of parameters yield to a
distribution similar to that shown in the upper panel of Fig.~\ref{color}.  A
somewhat sharper bimodality can be obtained only in models with high values of
the mass loss rate parameter $\eta$.
The results shown in Fig.~\ref{color} explain that our model predicts a lower
(with respect to DLB07) abundance of red ($B-V > 0.8$), faint galaxies because
the new feedback scheme leads to a steeper colour-magnitude relation.

\section{Consequences for the chemical evolution of galaxies}
\label{chemic}

In this Section we analyse how the stellar and gas metallicities of our model
galaxies relate to their mass or luminosity, and compare our results to
different observational measurements.  The ability of a galaxy formation model
to reproduce the observed correlations provides important constraints on how
well the circulation of baryons in the different components is modelled.  

As explained in \citet{kauffmann1996}, the use of a surface density threshold
for the star formation naturally reproduces the observed trend of the gas
fraction as a function of galaxy luminosity.  Our results are very similar to
those shown in \citet{delucia2004} and \citet{croton2006} so we do not repeat
them here.

\subsection{The metallicities of stars}
\label{zs}

\begin{figure}
\centering
\includegraphics[width=8.4cm]{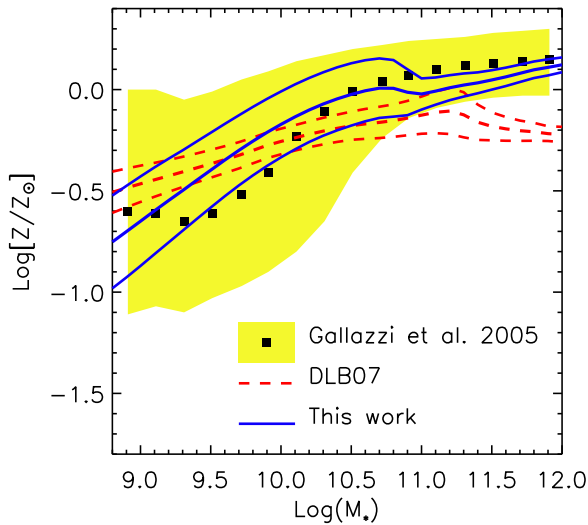}
\caption{Stellar metallicity as a function of stellar mass. The large dots and
  the shaded region represent the median and the 16th and 84th percentiles of
  the distribution measured for a sample of SDSS galaxies by
  \citet{gallazzi2005}.  The solid and dashed lines show the same quantities
  for the model presented here and for the model of DLB07 respectively.}
\label{zstars}
\end{figure}

Fig. \ref{zstars} shows our results for the metallicity of the stellar
component of galaxies and compares them to the observational measurements by
\citet{gallazzi2005} and to the model results by DLB07.  All model galaxies
have been used here.  \citet{gallazzi2005} analyse about 44000 galaxies from
the SDSS DR2 and find that the stellar metallicity increases with stellar mass
up to $M_{\star} \sim 10^{10.5}$ \hm M$_{\sun}$. For larger masses, the metallicity distribution flattens and saturates to a super--solar value.  The stellar metallicity also flattens for $M_{\star} \lesssim 10^{9.5}$ \hm M$_{\sun}$. Our model predicts a trend roughly similar to the observational data, although the distribution of the simulated metallicities shows a smaller scatter. The model discussed in DLB07 exhibits an even smaller scatter and a shallower relation that appears to be in less good agreement with observational results. None of the two models exhibits the observed flattening for $M_{\star} < 10^{9.5}$ \hm M$_{\sun}$.
Our predictions also broadly agree with those of \citet{monaco2007}.

The change in slope of the metallicity-mass relation in our model is a consequence of the feedback scheme.  As we will show in Section \ref{enrich}, galaxies in haloes with $M\vir < 10^{12}$ \hm M$_{\sun}$ eject a larger fraction of their gas mass in winds and in the IGM than galaxies in larger haloes. This halo mass roughly corresponds to stellar masses of about $M_{\star} \sim 10^{10.5}$ \hm M$_{\sun}$. The high efficiency of mass ejection in small haloes strongly suppresses the formation of stars in these galaxies and therefore affects the amount of metals that can be produced and locked into their stellar component, ultimately shaping the distribution of the stellar metallicities.  We have experimented with different sets of parameters and verified that the flattening of the slope always appears where the efficiency of winds in ejecting gas from galaxies drops.

\subsection{Luminosity--gas metallicity relationships}
\label{zism}

\begin{figure}
\centering
\includegraphics[width=8.4cm]{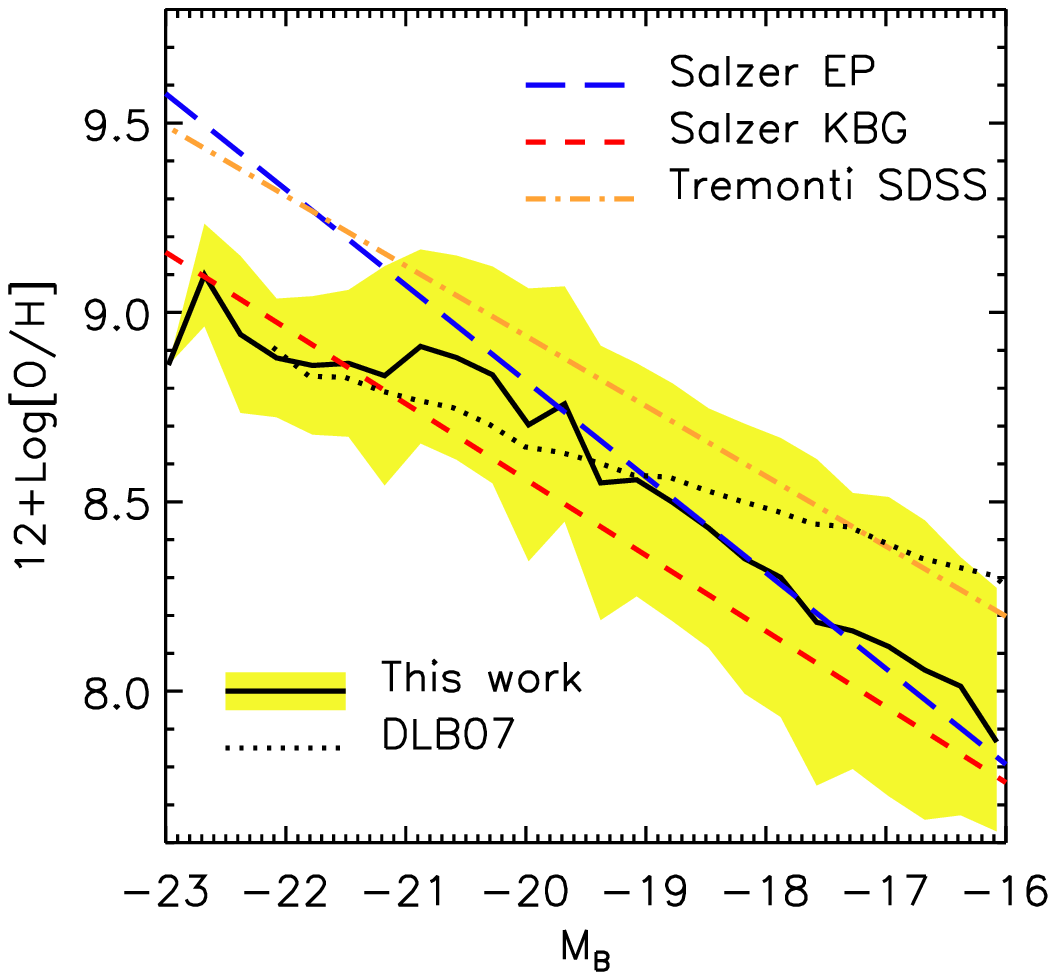}
\includegraphics[width=8.4cm]{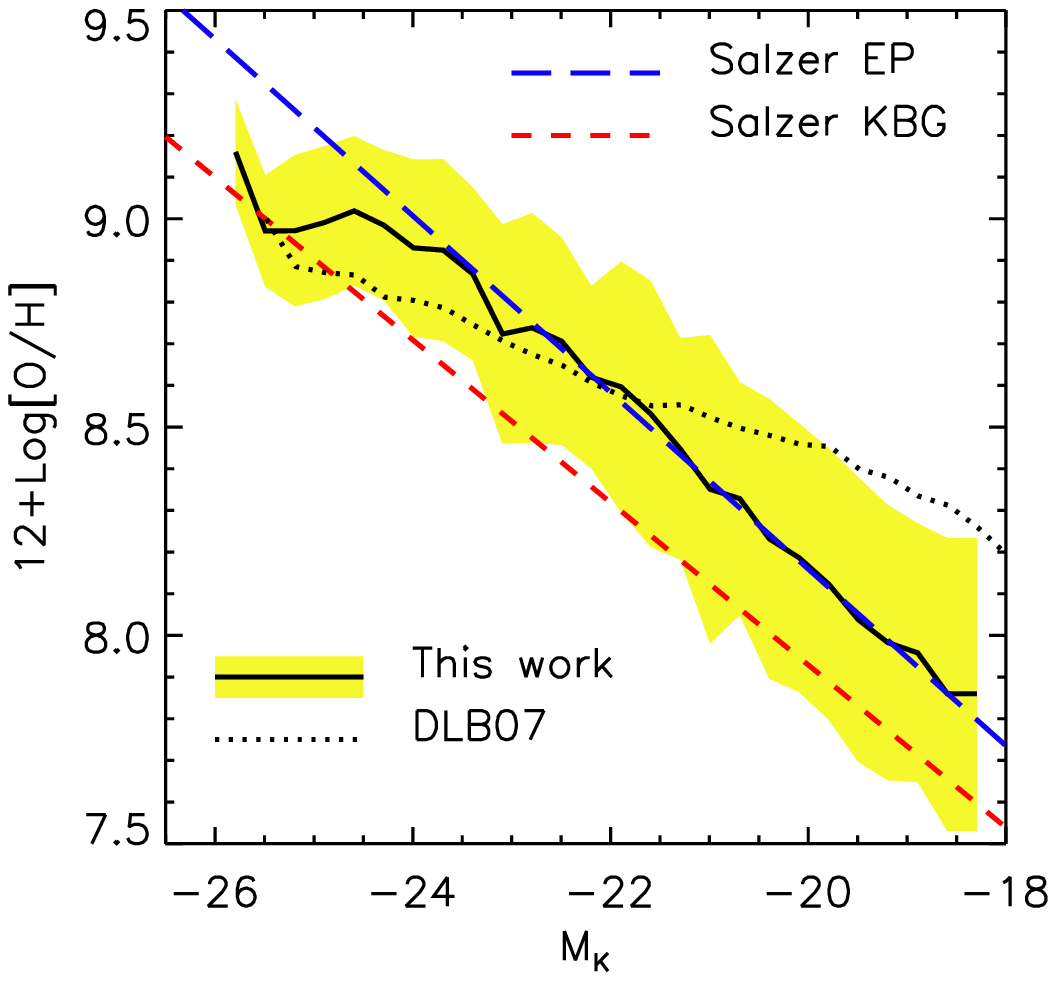}
\caption{Gas metallicity as a function of the magnitude in B--band (upper
  panel) and in the K--band (lower panel). The black solid line shows the
  median of the distribution for model galaxies.  The yellow marks the region
  including the 16th and 84th percentiles of the distribution.  Observational
  determinations from \citet{salzer2005} (using two different calibrations) and
  \citet{tremonti2004} are shown as dashed and dot-dashed lines respectively.}
\label{magz}
\end{figure}

Fig. \ref{magz} shows the optical luminosity--gas metallicity relationships in
B--band and K--band and compares the model results with fits to observational
data from \citet{salzer2005} and \citet{tremonti2004}.  \citet{tremonti2004}
estimate the oxygen abundance in SDSS galaxies by fitting simultaneously the
most prominent emission lines with a model for integrated galaxy spectra
\citep{charlot2001}.  \citet{salzer2005} use a sample of several hundred
galaxies from the Kitt Peak National Observatory International Spectroscopic
Survey (KISS, \citealt{salzer2000}). They use two different metallicity
calibrations: from Edmunds \& Pagel (1984, hereafter EP) and from Kennicutt et
al. (2003,  hereafter KBG).  As can be seen from Fig.~\ref{magz}, different
metallicity calibrators give slightly different estimates of the oxygen
abundance in nebular emission lines.  Results from our model appear to be in
very good agreement with the EP fit by \citet{salzer2005} for $M_{\rm B} < -20$
and $M_{\rm K} < -24$.  At the brightest magnitudes, the luminosity--gas
metallicity relation changes slope and model results appear to be closer to the
fit corresponding to the KBG calibrator.  The estimate of \citet{tremonti2004}
intercepts the upper end of the distribution of the numerical results in
B--band.

To estimate the oxygen abundance of model galaxies, we
assume a solar oxygen abundance of $\oxy = 8.76$. The latest determinations of
the solar oxygen abundance range from $\oxy = 8.66$ as estimated by
\citet{asplund2006} using photospheric abundances, to $\oxy = 8.86$ as
estimated by \citet{dp2006} on the basis of helioseismology results.  The
higher value of $\oxy$ would bring our results in better agreement with the fit
by \citet{tremonti2004}, while the lower value would give a better agreement
with the KBG fit.  Independently of the normalization adopted, however, our
predictions nicely match the observed slope of the luminosity--metallicity
relationship in both B--band and K--band.  Different combinations of model
parameters shift the relation upwards or downwards by a few percent, but do not
affect the slope or the scatter of the distribution.  The dotted lines in
Fig.~\ref{magz} show the corresponding relationships from the model of
DLB07.  When compared to our model results, this model predicts a shallower
relation in the B--band and a non--linear relation in the K--band.  Faint
galaxies with $M_{\rm B} > -18$ and $M_{\rm K} > -20$ show the largest
deviations from the observed metallicities. The model of DLB07 also features a smaller scatter than our model.

Overall, the good agreement between the observed stellar and nebular metallicities and the predictions of our model is an indication that the dynamical feedback scheme for galactic winds is well integrated with the chemical evolution of galaxies. On the other hand, Figs. \ref{zstars} and \ref{magz} indicate that the model of DLB07 may significantly overestimate the metallicities of faint galaxies, indicating perhaps an insufficient ejection of metals in these galaxies.

\section{The recycling of mass and metals in galactic winds}
\label{enrich}

In this Section we focus our analysis on the recycling of metals in feedback processes. We trace the history of the deposition of metals into the IGM through cosmic time and investigate which haloes are the main sources of the metal enrichment of the IGM. Throughout, we highlight the main differences between our feedback scheme and that of DLB07.

Feedback processes, and supernova feedback in particular, determine the recycling of gas between episodes of star formation. Only a fraction of the galaxies in the simulation release enough energy in supernovae explosions to blow winds to distances comparable or larger than the virial radius of the halo. As a consequence, not all winds can deposit mass and metals into the IGM.
Most winds start blowing with such a little initial energy, compared to the energy required to overcome the gravitational and pressure forces of the halo, that they collapse within a few time--steps. The role of these winds is important in regulating the star formation history of the central galaxies, because they determine the removal of the material reheated by supernovae explosions and its deposition into the hot halo gas.  

We assume that winds escape haloes when they reach the Hubble flow. When this happens, the wind mass and metals are deposited into the IGM and do not participate anymore in the evolution of galaxies in the halo. When winds reach the Hubble flow, the density field in which they expand cannot be estimated analytically, unless the real 3--dimensional distribution of matter in the simulation is taken into account. We are therefore unable to trace the spatial distribution of these metals. In addition, our model neglects the possibility that metals mixed with the intergalactic gas can be incorporated at later epochs by nearby haloes or by the same parent halo accreting mass from a larger infall region.

\begin{figure}
\centering
\includegraphics[width=8.4cm]{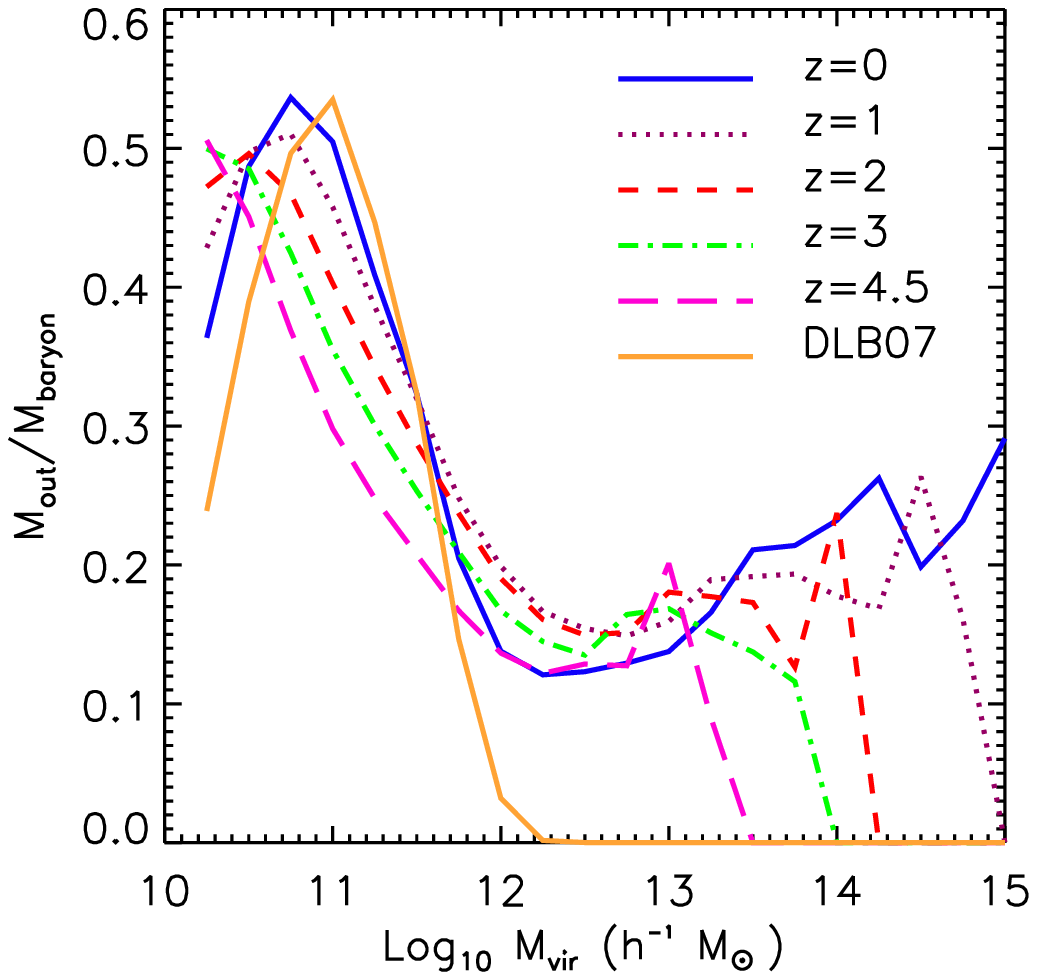}
\includegraphics[width=8.4cm]{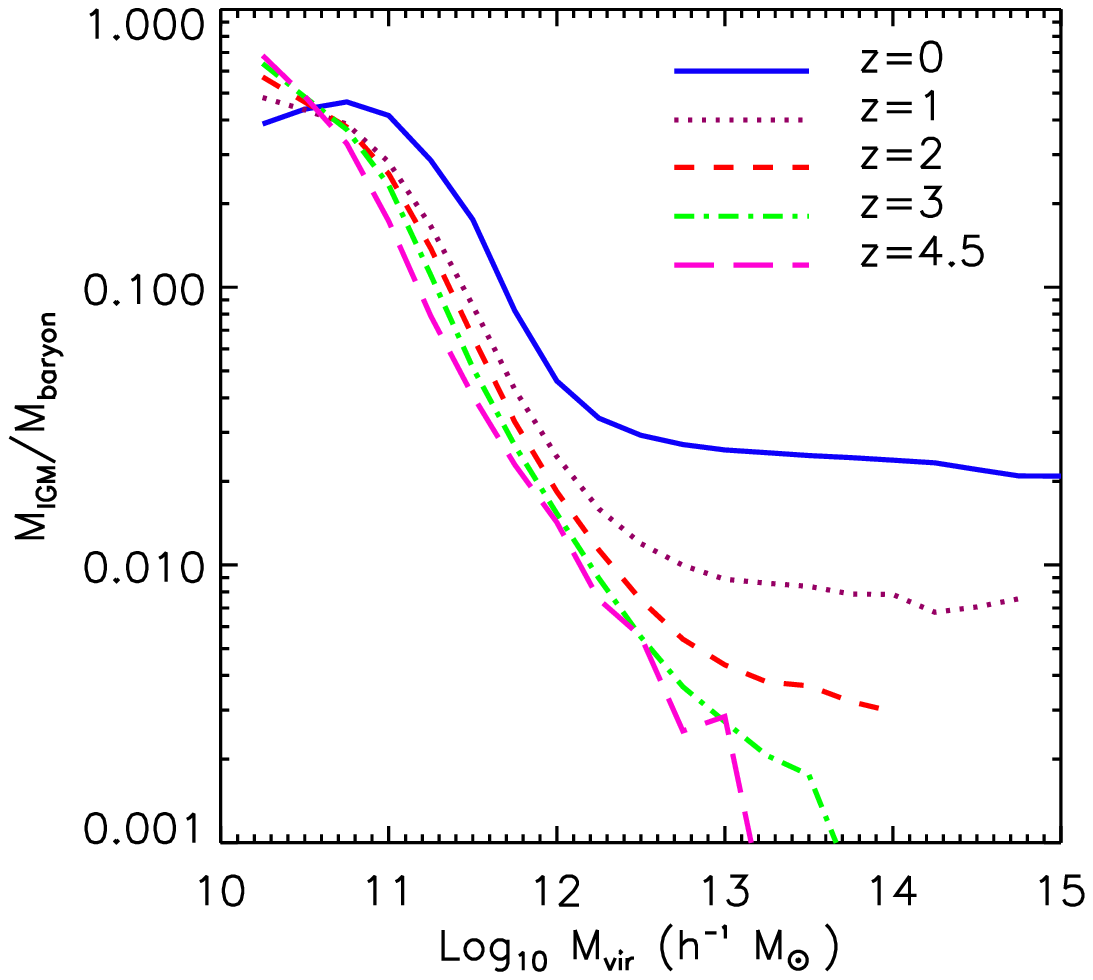}
\caption{Average fraction of ejected mass $M_{\textrm{out}}$ (upper panel) and of mass deposited into the IGM $M_{\textrm{IGM}}$ (lower panel), as a function of halo virial mass. The ejected mass $M_{\textrm{out}}$ is the sum of the mass in winds, plus the mass deposited into the IGM $M_{\textrm{IGM}}$. The solid yellow line in the upper panel shows the fraction of ejected mass in the DLB07 model at $z=0$. Results are shown at $z=0, 1, 2, 3$ and 4.5.} 
\label{gasout}
\end{figure}

\begin{figure}
\centering
\includegraphics[width=8.4cm]{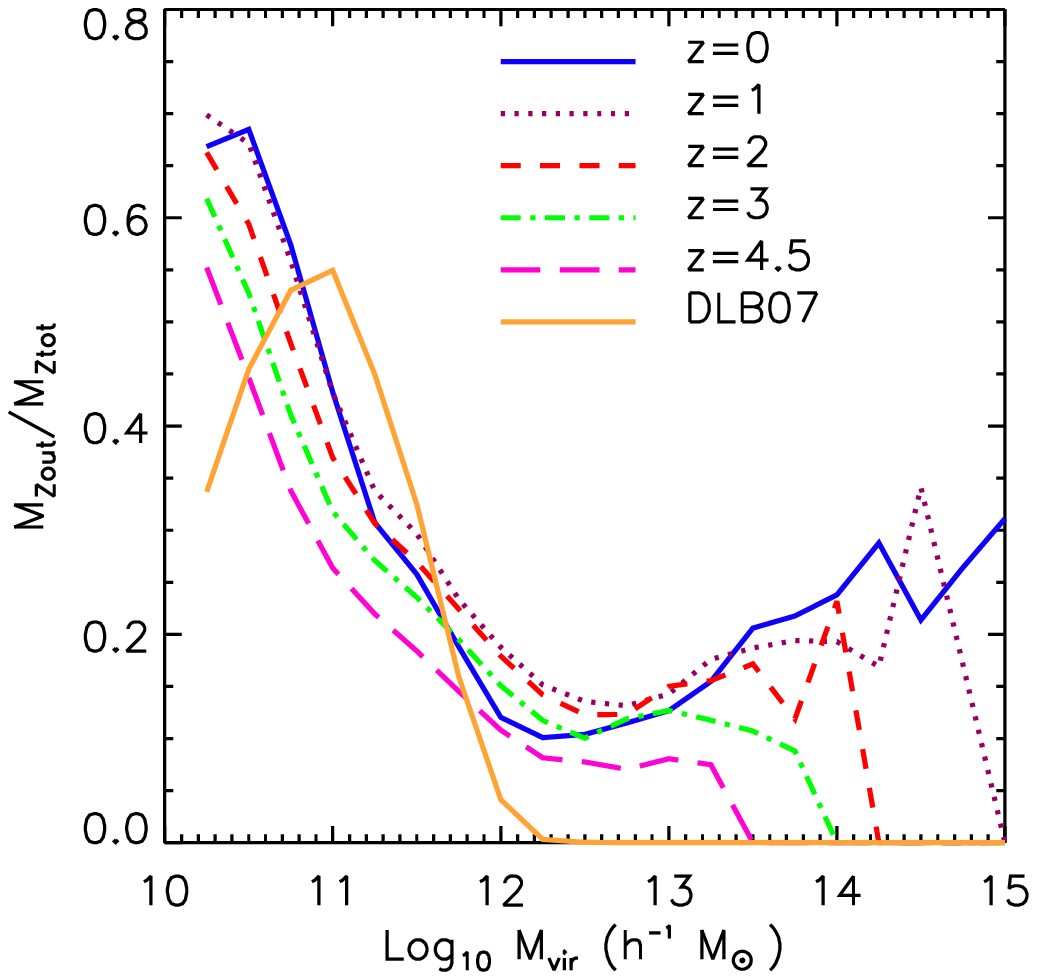}
\includegraphics[width=8.4cm]{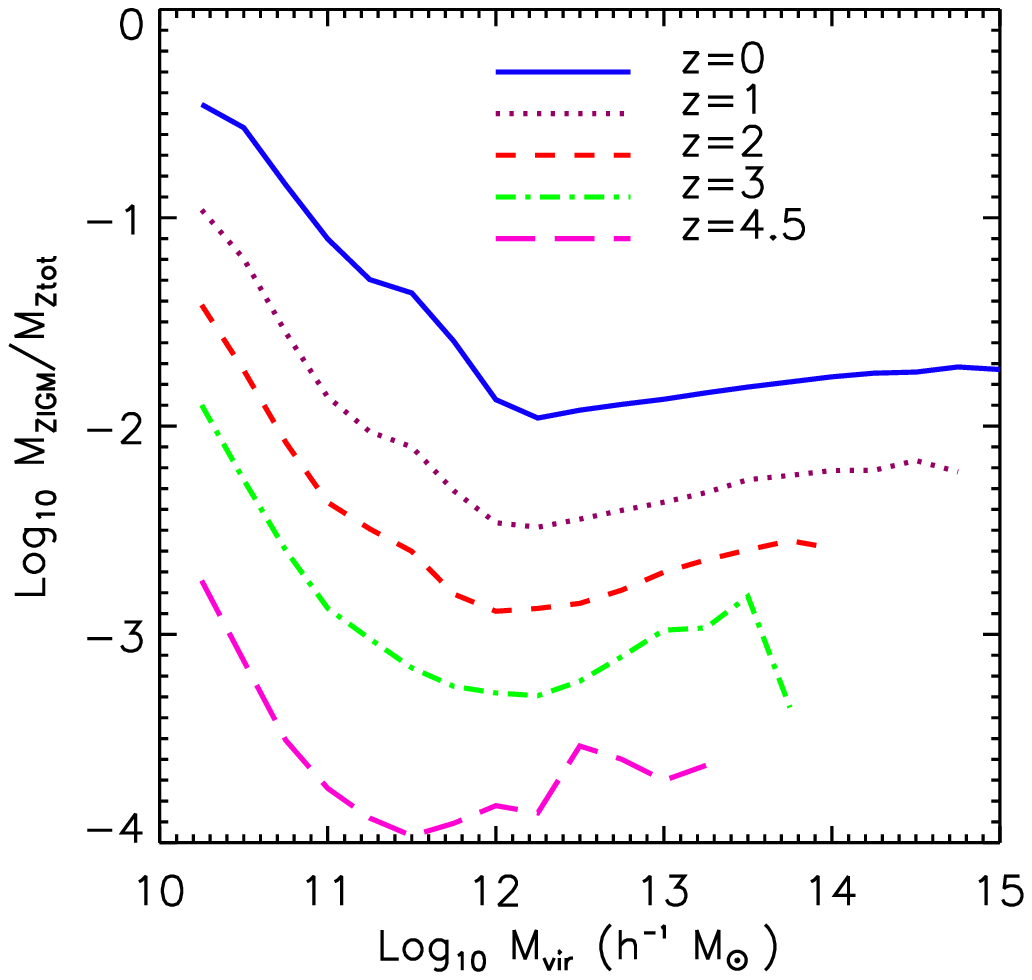}
\caption{Average fraction of ejected metal mass $M_{\textrm{Zout}}$ (upper panel) and of metal mass deposited into the IGM $M_{\textrm{ZIGM}}$ (lower panel), as a function of halo virial mass. The ejected metals mass $M_{\textrm{Zout}}$ is the sum of the metals in winds, plus the metals deposited into the IGM $M_{\textrm{ZIGM}}$. $M_{\textrm{Ztot}}$ is the total amount of metals in haloes. The solid yellow line shows the fraction of ejected metal mass in the DLB07 model at $z=0$. Results are shown at $z=0, 1, 2, 3$ and 4.5.} 
\label{metalsout}
\end{figure}

Figs. \ref{gasout} and \ref{metalsout} show how efficiently mass and metals, respectively, are ejected by winds and deposited into the IGM.
The upper panel of Fig. \ref{gasout} shows the average fraction of mass ejected by winds $M_{\textrm{out}}$, defined as the sum of the mass currently in winds, plus the mass deposited into the IGM, $M_{\textrm{IGM}}$, over time. Similarly, the upper panel of Fig. \ref{metalsout} shows the average fraction of the metal mass ejected by winds $M_{\textrm{Zout}}$, again defined as the sum of the metal mass currently in winds, plus the metals deposited into the IGM, $M_{\textrm{ZIGM}}$, over time.
$M_{\textrm{baryon}}$ is the total baryonic mass of the halo, while  $M_{\textrm{Ztot}}$ is the total metal mass in haloes and includes the metals in the gaseous phases and the metals locked in stars.
The solid yellow lines are the fractions of ejected mass and metals in the DLB07 model at $z=0$, respectively. The lower panel of Fig. \ref{gasout} shows the average fraction of mass deposited into the IGM $M_{\textrm{IGM}}$, while the lower panel of Fig. \ref{metalsout} the average fraction of metal mass deposited into the IGM, $M_{\textrm{ZIGM}}$. Results are presented as a function of halo virial mass for $z=0, 1, 2, 3$ and 4.5.

The mass and metal ejection are most efficient in haloes with $M\vir < 10^{12}$ \hm M$_{\sun}$. Winds from these haloes can eject up to $60$ per cent of the baryons in the halo, of which more than half may have been deposited into the IGM. For haloes with $10^{11}$ \hm M$_{\sun} < M\vir < 10^{12}$ \hm M$_{\sun}$, the DLB07 model predicts a mass ejection efficiency comparable to ours, but a significantly higher metal ejection efficiency. 
Larger haloes corresponding to groups and clusters can deposit at most a few percent of their mass and metals into the IGM, in agreement with the widely used assumption that the chemical evolution of clusters can be described at first order by a closed--box model.
We will discuss the behaviour of the distribution of the ejected mass for $M\vir > 10^{12}$ \hm M$_{\sun}$ at the end of this Section.

At $z=0$, our model predicts a mass ejection efficiency significantly larger 
than that in the model used in DLB07 for haloes with with 
$M\vir \lesssim 10^{11}$ \hm M$_{\sun}$. This difference is responsible for 
the lower abundance of dwarf galaxies predicted by our model. The same haloes eject in our model on average twice the mass of metals ejected in the DLB07 model. This explains why our model produces on average dwarf galaxies with lower gas  and stellar metallicities than the DLB07 model, as shown in Figs. \ref{zstars} and \ref{magz}. We note however that haloes in this mass range contain typically less than a few hundreds particles and might therefore be affected by resolution. Results in this mass range should be interpreted with caution. The mass and metal ejection efficiencies are only mildly dependent on redshift. However, metals are deposited into the IGM more efficiently with decreasing redshift.

\begin{figure}
\centering
\includegraphics[width=8.4cm]{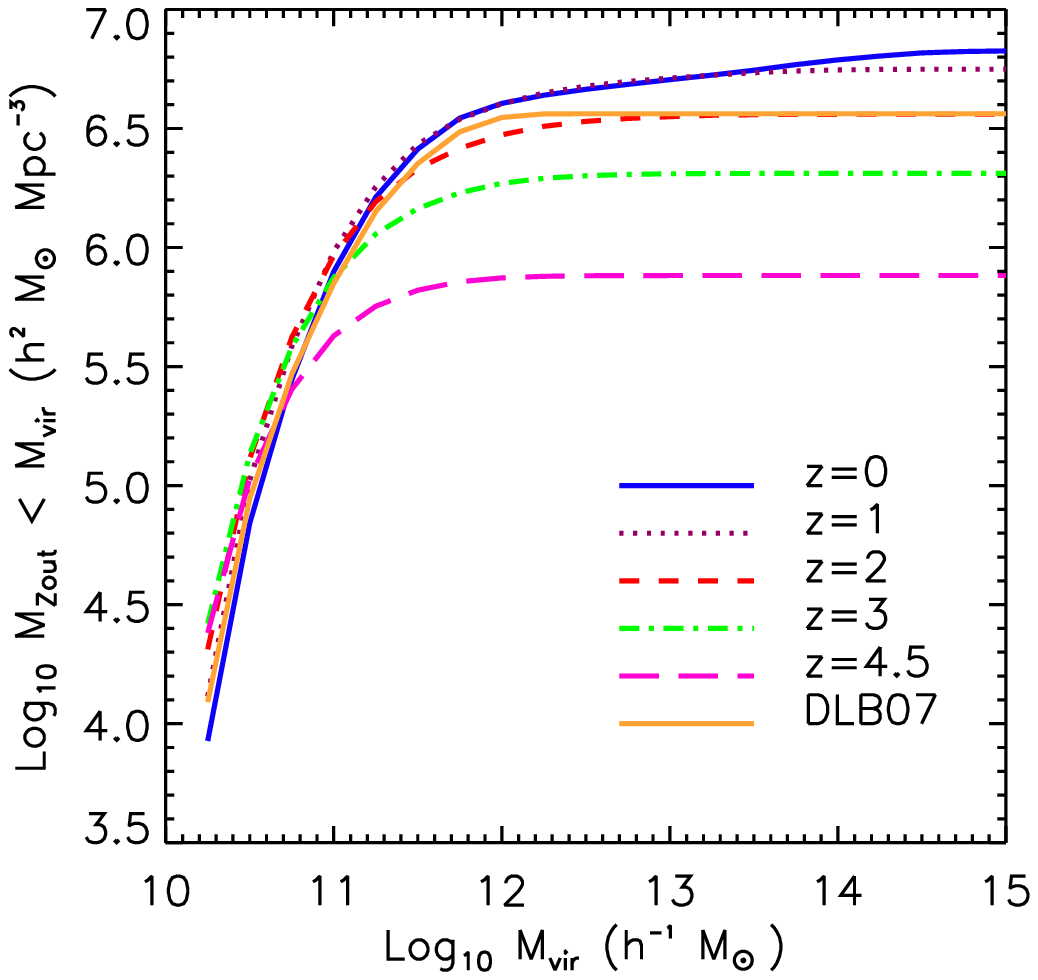}
\includegraphics[width=8.4cm]{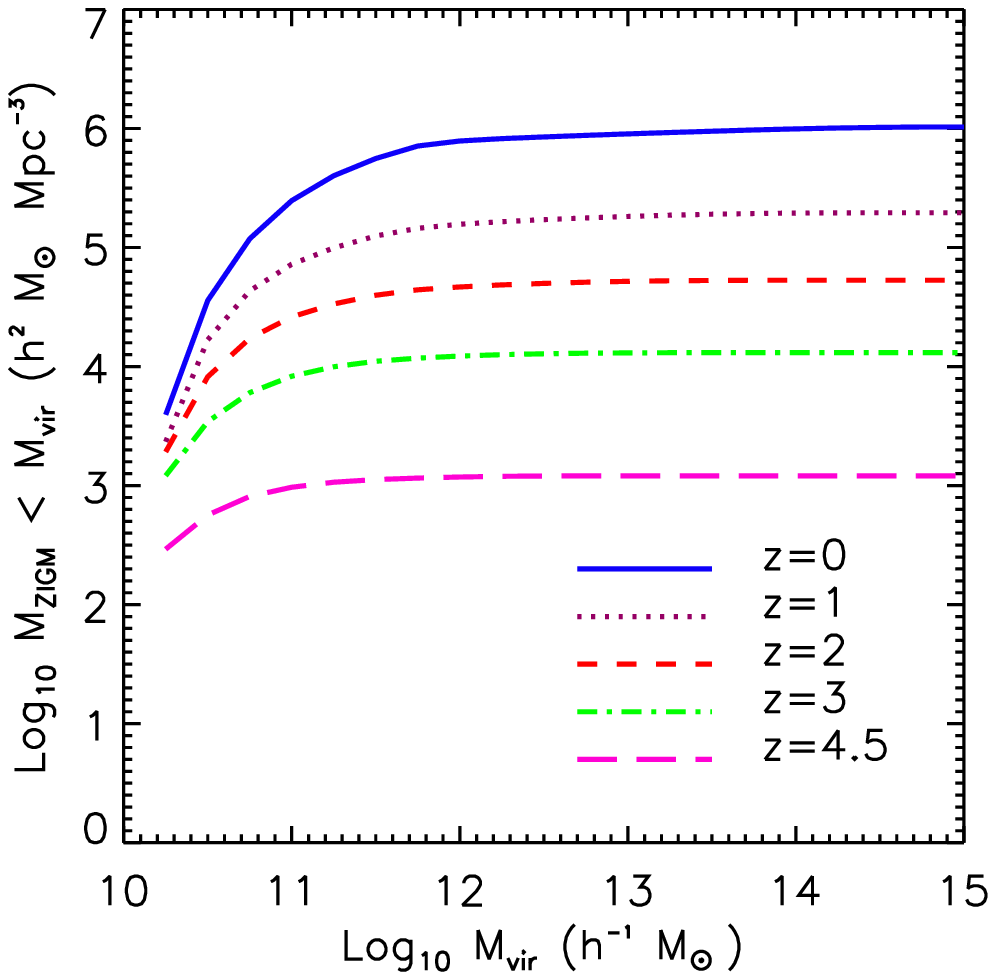}
\caption{Cumulative distribution of the ejected metal mass $M_{\textrm{Zout}}$ (upper panel) and of the metal mass deposited into the IGM $M_{\textrm{ZIGM}}$ (lower panel) per unit of comoving volume, as a function of halo virial mass. The solid yellow line in the upper panel shows the ejected mass in the DLB07 model at $z=0$. Results are shown at $z=0, 1, 2, 3$ and 4.5.} 
\label{cumwinds}
\end{figure}

Fig. \ref{cumwinds} shows the cumulative distribution of the total metal mass density ejected by winds (upper panel) and of the metal mass density deposited into the IGM (lower panel) as a function of halo virial mass. Results are presented at $z=0, 1, 2, 3$ and 4.5.
The solid yellow line shows the results of DLB07 at $z=0$.
$M_{\textrm{ZIGM}}$ and $M_{\textrm{Ztot}}$ are calculated as in Fig. \ref{metalsout} and the mass density is expressed per unit of comoving volume.
Fig. \ref{cumwinds} confirms the idea that the largest contribution to the ejection of metals and to the metal enrichment of the IGM comes from galaxies in haloes with $M\vir < 10^{12}$ \hm M$_{\sun}$.
At $z=0$, the distribution of the ejected metals in the DLB07 model closely follows that predicted by our model for haloes with $M\vir < 10^{12}$ \hm M$_{\sun}$.
Fig. \ref{cumwinds} also suggests that, while at $z=0$ most of the metals in the IGM are not gravitationally bound to haloes, at $z=2-3$ most metals are still in the process of being ejected by winds and have not yet reached the Hubble flow.
Our findings are broadly consistent with those of \citet{bouche2007}. \citet{bouche2007} investigated the metal budget in the IGM at $z\sim 2.5$ and $z\sim 0$ and found that galaxies with $L_{\textrm{B}} < 0.3 L_{\textrm{B}}^*$ are responsible for about 90 per cent of the metal enrichment of the IGM. Our results for the mass density of ejected metals support their claim and we predict a similar amount of metals ejected by galaxies per unit volume, as does the DLB07 model.

According to the upper panels of Figs. \ref{gasout} and \ref{metalsout}, haloes with $M\vir > 10^{12}$ \hm M$_{\sun}$ can transfer up to 25\% of their mass into a wind, before this is incorporated into the hot gas. This is an artifact of our feedback scheme. When a wind attempts to blow out of a massive halo, its initial velocity and the halo central gas density are high; this means that a much larger mass is accreted onto the wind from the halo gas than from the material outflowing from the star--forming galaxies. This effect becomes stronger with increasing halo mass, hence the trend of increasing ejection efficiency for $M\vir > 10^{12}$ \hm M$_{\sun}$.
Since winds in massive haloes always collapse after a few timesteps, this feature simply describes the deposition of mass and metals into the hot cluster gas. The mass and metals ejected by haloes with $M\vir > 10^{12}$ \hm M$_{\sun}$ are therefore not relevant for the global mass and metal budget.
This demonstrates to what extent our feedback scheme is different from that of DLB07.
This mass transfer does not require an intermediate stage in the DLB07 model (see \citealt{croton2006} for details). In DLB07, the hot gas is ejected only if the energy transferred to the hot gas by supernova explosions is larger than the increase of the thermal energy of the halo due to the deposition of the reheated mass. Since this never happens in haloes with $M\vir \gtrsim 10^{12}$ \hm M$_{\sun}$, the hot gas is never removed from the halo. Hence the difference in the amount of gas being recycled in these haloes in the two models.
The different treatment of feedback in haloes with $M\vir > 10^{12}$ \hm M$_{\sun}$ also explains why the model of DLB07 produces only red galaxies at the highest stellar masses, while our model still predicts a few blue, late type galaxies (Fig. \ref{fracs}). The continuous recycling of gas in groups with $M\vir \sim 10^{12}-10^{15}$ \hm M$_{\sun}$ may favour the cooling of gas onto the central galaxies, which may accrete a sufficient amount of gas to form a disk.

To verify that our findings are genuine and do not critically depend on the speficic values of the model parameters, we modify the power law index $\alpha_{\textrm{w}}$ in eq. (\ref{massloss}) and (\ref{windspeed}) and we consider models with $\alpha_{\textrm{w}} = 0, 1$ and $1.5$ (for consistency, the value of the wind initial velocity $V_{\textrm{w}}$ is also adjusted to keep the total wind energy roughly constant). The smaller the value of $\alpha_{\textrm{w}}$, the less mass is lost by smaller galaxies and the larger the mass injected in winds by larger galaxies. However, we find that no matter the value of $\alpha_{\textrm{w}}$, haloes with $M\vir < 10^{12}$ \hm M$_{\sun}$ are always the most efficient to deposit metals into the IGM, while larger haloes do not lose more than a few percent of their metals. This confirms that the dynamical feedback scheme we present here describes the evolution of galactic winds and their effects on galaxies, independently of the actual values of the model parameters.

\section{Conclusions}
\label{summary}

In this work, we have presented a new feedback scheme which is based on a
dynamical treatment of the evolution of galactic winds in a cosmological
context.  The scheme has been incorporated in the Munich semi--analytic model
for galaxy formation described in \citet{delucia2007} and
\citet{croton2006} and replaces previous implementations of supernova
feedback, but does not alter the AGN feedback model. In our scheme, the
ejection and recycling of gas and metals are treated self--consistently and are
governed by the dynamical evolution of winds, rather than by empirical
prescriptions.

Our model contains a number of assumptions, as we have discussed above. These are:
i) winds are described as spherically symmetric blastwaves, as in \citet{serena}.
ii) The dynamics of winds is described as a two-stage process, e.g. a pressure--driven, adiabatic stage, followed by a momentum--driven snowplough. An intermediate pressure--driven snowplough stage is not included in our treatment of winds.
iii) Winds sweep up all the gas they encounter along their path. This situation corresponds to an entrainment fraction parameter $\varepsilon =1$ in the model of \citet{serena}.
iv) Winds are linearly added together when haloes merge, without being destroyed in the process.
v) When winds collapse, the wind mass is instantaneously reincorporated into the hot halo gas.
vi) We do not track the injection of energy into the hot halo gas when winds collapse and their mass is reincorporated into the halo hot gas.
vii) When winds reach the Hubble flow, their mass is permanently deposited into the IGM.
viii) Winds are powered by the star formation taking place in all galaxies in the halo, including satellites.
ix) The wind mass loss rate $\dot{M}_{\textrm{w}}$ and the wind initial velocity $v_{\textrm{wo}}$ are described by Eqs. \ref{massloss} and \ref{windspeed}, respectively.

Our results demonstrate that our dynamical scheme is a good alternative to empirical recipes for feedback and in particular it is able to reproduce the observed properties of dwarf galaxies with higher accuracy than previous models.
Our main results can be summarized as follows:

\begin{enumerate}
\item The predictions for the the K--band and \bj--band luminosity functions and for the stellar mass function are in good agreement with observations. In particular, the number density of dwarfs is reproduced with greater accuracy than in the model presented in DLB07. 
  
\item The stellar mass--stellar metallicity and the luminosity--gas metallicity
  relationships are in good agreement with observations and significantly
  improve previous results. Again, the improvement is significant for dwarf
  galaxies.
\end{enumerate}

These results demonstrate that in our model the treatment of feedback in dwarf galaxies is significantly improved with respect to previous implementations.
We argue that the main reason for this improvement is the possibility of
estimating the amount of mass and metals that a halo can permanently lose and
deposit into the IGM.  This feature determines a stronger quenching of star
formation in small haloes and therefore the suppression of the faint end of the
stellar mass and luminosity functions of galaxies, and the shape of the
mass--stellar metallicity and of the luminosity--gas metallicity relationships.
  
Our model, however, also exhibits some weaknesses. Firstly, there is a tendency
to overestimate the number of bright galaxies in the LFs, because of the
continuous recycling of gas when winds recollapse in massive haloes.  A
possible solution could be an increase in the efficiency of AGN feedback (about
20--30 per cent more intense than suggested by \citealt{croton2006}), but this
would negatively affect the abundance of galaxies at the knee of the LFs.  Our
model does not solve the problem of the abundance of $M^{\ast}$ galaxies at
$z=0$, which continues to be underestimated (see also \citealt{bower2006}).
Finally, perhaps the most severe shortcoming of the model is that it does not
correctly reproduce the colour distribution of galaxies.  In particular, our
model results do not exhibit the sharp colour bimodality observed for galaxies
in the local Universe.  This feature does not depend on the specific values of
the model parameters and it can be alleviated only by assuming very
inefficient feedback. However, models with inefficient feedback do not correctly reproduce the abundance of dwarf galaxies, nor their metallicities.

In the second part of our paper, we have investigated the efficiency of the
mass and metal injection in winds and their deposition into the IGM.  We have shown that in our model, galaxies in haloes with $M\vir \lesssim 10^{12}$ \hm M$_{\sun}$ are responsible for most of the metal enrichment of the IGM, being the only objects that can blow winds out of their gravitational potentials.  Larger haloes like groups and clusters lose only a small fraction of their mass and metals, and this occurs before most of the halo mass is accreted.

Our feedback model contains the same number of free parameters (i.e. 3) as the
supernova feedback scheme of \citet{croton2006}, once the dependence of the ejected mass on the halo virial velocity in \citet{croton2006}, which corresponds to our parameter $\alpha_{\textrm{w}}$, is taken into account.  The numerical calculation of the evolution of winds is only slightly more time--consuming than previous feedback schemes. However, it does require about 10--15 per cent additional memory to handle the wind variables and to store them on disk, if desired.

\section*{Acknowledgments}
We would like to thank Jon Loveday and Bernard Pagel for useful discussions and Simon White for reading the manuscript. We also thank the referee, Pierluigi Monaco, for his insightful suggestions that helped to improve the quality of the paper.
SB is supported by PPARC.
The Millennium simulation was carried out by the Virgo Consortium at the Max Planck Society in Garching.
Data on the galaxy population produced by this model, as well as on the parent halo population, are publicly available at http://www.mpa-garching.mpg.de/millennium/.

\bsp

\label{lastpage}

\end{document}